# Orbital angular momentum multiplication in plasmonic vortex cavities

G. Spektor[1,2,3], E. Prinz[4], M. Hartelt[4], A.K. Mahro[4], M. Aeschlimann[4] and M. Orenstein[1].

[1]*Department of Electrical Engineering, Technion - Israel Institute of Technology, 32000 Haifa, Israel.*
[2]*Associate of the National Institute of Standards and Technology, Time and Frequency Division, Boulder, Colorado 80305, USA (curr. address)*
[3]*Department of Physics, University of Colorado, Boulder, Colorado 80309, USA (curr. address)*
[4]*Department of Physics and Research Center OPTIMAS, University of Kaiserslautern, Erwin Schroedinger Strasse 46, 67663 Kaiserslautern, Germany.*

spektorg@campus.technion.ac.il

**Abstract:** Orbital angular momentum of light is a core feature in photonics. Its confinement to surfaces using plasmonics has unlocked many phenomena and potential applications. Here we introduce the reflection from structural boundaries as a new degree of freedom to generate and control plasmonic orbital angular momentum. We experimentally demonstrate plasmonic vortex cavities, generating a succession of vortex pulses with increasing topological charge as a function of time. We track the spatio-temporal dynamics of these angularly decelerating plasmon pulse train within the cavities for over 300 femtoseconds using time-resolved Photoemission Electron Microscopy, showing that the angular momentum grows by multiples of the chiral order of the cavity. The introduction of this degree of freedom to tame orbital angular momentum delivered by plasmonic vortices, could miniaturize pump-probe-like quantum initialization schemes, increase the torque exerted by plasmonic tweezers and potentially achieve vortex lattice cavities with dynamically evolving topology.

Orbital angular momentum (OAM) of light has been of wide interest in recent years [1–6], laying the foundation for promising applications such as reshaping light-matter interactions [7,8] optical tweezing [9–12], optical communications [13–15] and quantum information processing [16]. The confinement of OAM to a surface enabled access and control to new physical phenomena [17,18] including linear spin-orbit conversion [19–21], orbit-orbit conversion [22] as well as nonlinear spin-orbit mixing [23]. Recent experimental access to the sub-cycle surface plasmon polaritons (SPP) dynamics [24–26] enabled the study of the dynamical evolution of a vortex [27] and direct determination of its angular momentum, as well as revealed the dynamics of plasmonic nano-focusing [28,29], thus raising the possibility of temporal OAM control. In parallel, the ability to generate vortex lattices [30–33] and the emergence of their topological properties [34,35] are of great current interest to the scientific community [36]. Revealing and taming new degrees of freedom for generating and controlling OAM is thus key to the advancement of the field.

The generation and shaping of plasmonic fields are typically achieved by engraving boundaries in a metal and exciting them with a certain illumination [20,21,24,25,28,29,34,35,37–44]. Be it continuous engravings or local geometries forming meta-slits [42,45–49], it is generally widely assumed that the generated fields are set by the initial interaction with the static boundaries of the engraved structures. Even more so, in the case of plasmonic OAM generation and control, it has practically become a paradigm that the boundaries defined by a plasmonic vortex generator define a single vortex, well described by a single Bessel distribution [19–21,27,28,43].

In this work, we challenge this paradigm. We experimentally demonstrate that after the initial SPP vortex is excited by the boundaries, it subsequently interacts with the boundaries resulting in partial reflections. Intriguingly, we show that these reflections form successive, previously unobserved higher-order vortices, gaining additional angular momentum at each reflection (Fig. 1 (A-C)). We demonstrate that even a slit geometry has weak-yet-measurable contributions to these high order vortices. To increase the reflected power we construct a plasmonic vortex cavity with ultra-flat gold surfaces and golden wall mirrors created by template stripping [50–54]. Employing this cavity, (Fig. 1 (D-F)) we produce a train of plasmonic vortices with dynamically increasing large OAM and therefore decelerating rotation velocity (See movies in SI). Using a time-resolved pump-probe scheme (Fig. 1 (F)) in photoemission electron microscopy (TR-PEEM) [26] we resolve the complex dynamics within the decelerating cavities, separating the successive vortex pulses in time and space and quantifying the rules governing the OAM growth. The flatness of our surface provides low scattering losses, allowing us to record dynamics for the long duration of about 330 fs corresponding to ~100 microns of accumulated propagation inside the cavity.

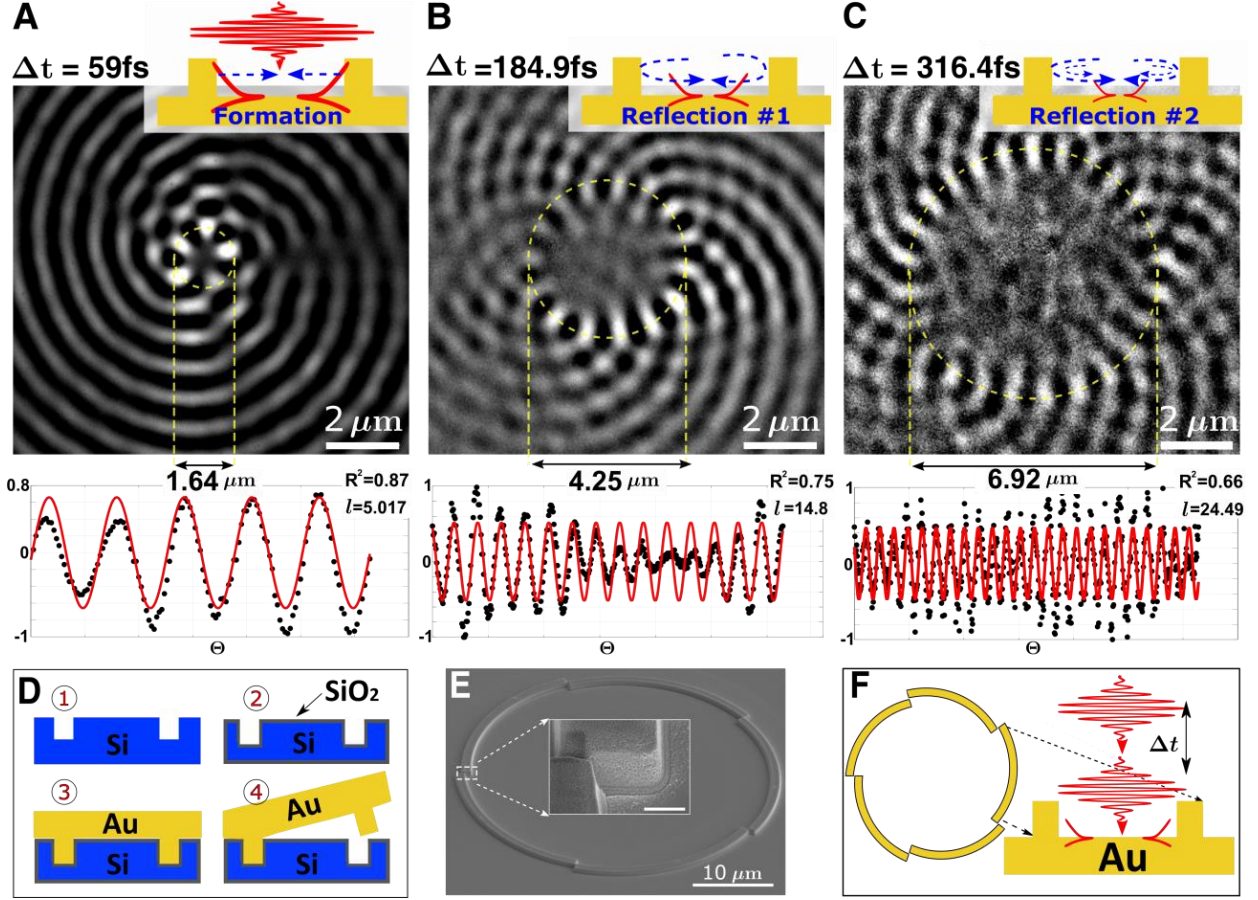

**Figure 1.** Experimental results (A-C) within a plasmonic vortex cavity of order m=5, showing the revolution stages [27] of the initial vortex of order $l$=5+1 (A) and subsequent first (B) and second (C) reflections of the wave packet, forming vortices of orders $l$=15+1 and $l$=25+1 respectively. The snapshots show 5, 15, and 25 azimuthal lobes imprinted into the photoemitted electrons due to the nonlinear mechanism of our technique [23]. The bottom insets show azimuthal and radial fitting of the diameter as well as the number of lobes in the main vortex signal. (D) Schematic of the template stripping process used to fabricate plasmonic vortex cavities with ridge mirrors; ① Focused Ion Beam milling of the inverse cavity in Silicon, ② Thermal oxide growth, ③ Au sputtering, and ④ Stripping. (E) SEM image of a template stripped vortex cavity of order $m$=5. The inset scale bar is 500nm. (F) Pump-probe setup schematic showing two ~23fs pulses with a center wavelength of 800nm (red) impinging on the sample at normal incidence with interferometrically-controlled time delay $\Delta t$ used in the TR-PEEM setup.

The formation of a plasmonic vortex is achieved when an azimuthally varying boundary in the form of an Archimedean spiral, $r(\theta) = r_i + \lambda_{spp} \cdot m\theta/2\pi$, is illuminated with properly polarized light [19,20]. Here $r$ is the distance of the boundary from the center, $r_i$ is the minimal distance, $\lambda_{spp}$ is the plasmonic wavelength, $m$ is the chiral order of the spiral structure, and $\theta$ is the azimuthal coordinate. In order to excite high-order vortices with an Archimedean spiral, a large $m$ is required. This causes significant propagation distance differences between plasmons originating from different locations along the slit, resulting in angularly dependent losses that distort the vortex. To mitigate this, the spiral is segmented [43] to the form $r_m(\theta) = r_i + \lambda_{spp} \cdot mod(m\theta, 2\pi)/2\pi$ resulting in a maximal path difference of $\lambda_{spp}$ within each segment, while still providing the necessary phase allowing the formation of high order vortices. In a typical setting, a spiral structure of order $m$ is illuminated by circularly polarized light with $\sigma = \pm 1$. One can describe a SPP pulse forming the vortex in the time domain by extending the schematic representation established by [43] (Fig. 2). The excitation of SPPs takes place at the boundaries while imprinting an azimuthal phase delay, that can be envisioned on an imaginary line equidistant from the center (Fig. 2 (A)). Then the still inward-propagating SPPs interfere with the already-outward propagating SPPs, forming the revolution stage of the vortex (Fig. 2 (B)). Subsequently, the vortex dissolves when the wave packet becomes predominantly outward-propagating, leaving the center of the structure while maintaining its azimuthal phase (Fig. 2 (C)). Indeed, the experiments reported to-date show great agreement with this description. It is accepted that in a sufficiently large structure, the electric field components at the vortex revolution stage/phase are very well approximated by a Bessel functional form given by, $E_z \approx J_l(k_{spp}r) \times exp(il\theta)$, $E_r \approx J_l'(k_{spp}r) \times exp(il\theta)$ and $E_\theta \approx J_l(k_{spp}r) \times exp(il\theta)$ where $J_l$ is the Bessel function of order $l$, $J_l'$ is its first derivative, $r, \theta, z$ are the cylindrical coordinates, $k_{spp}$ is the plasmonic wave number, and $i = \sqrt{-1}$. The vortex is then said to be of order $l$ and in the case of circularly polarized illumination $l = m + \sigma$. Historically the out-of-plane field component is the one mainly discussed in the literature due to its accessibility using scanning near field optical microscopy techniques. For continuity, we follow this practice here as well.

When the outward-propagating SPPs meet the boundaries that launched them, they are partially back-reflected and go back towards the center of the structure (Fig. 2 (D)). Notably, at each azimuthal location, the SPPs traverse the region enclosed between the imaginary dashed circular line and the physical boundaries of the structure twice, once before the reflection and once after.

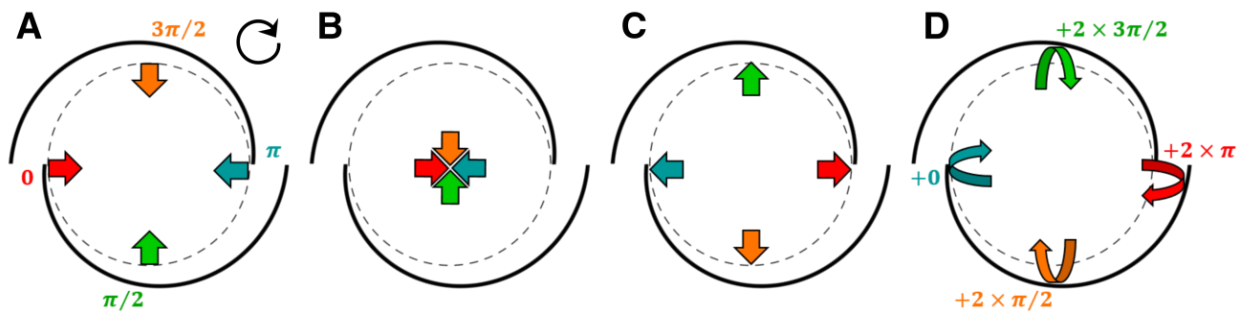

**Figure 2.** Schematic description of a round trip within a plasmonic vortex cavity. (A) A vortex generator of order $m=2$ illuminated by circular polarization (solid black arrow), generates counter-propagating SPPs with azimuthally varying phase delay (colored arrows) as depicted at the imaginary dashed circle concentric with the generator. (B) The plasmonic vortex is generated by the inward propagating SPP wave packets combining at the center. (C) After passing through the center, the outgoing wave packets keep the relative phase differences until they reach the imaginary dashed circle. (D) The SPPs propagate up to the cavity boundaries where they are partially reflected and propagate inward towards the center again; thus accumulating azimuthally varying relative phases corresponding to *twice* the geometric phase of the cavity. (Following the schematic representation established by [43]).

The reflected SPPs thus obtain an additional azimuthal phase equal to twice the order of the structure which results in the formation of a new vortex of the form

$$E_z \approx J_{l+2 \cdot m}(k_{spp} r) \cdot e^{-i(l+2m)\theta}.$$

This reflected vortex has different radial as well as azimuthal spatial distributions compared to the initial vortex and in a steady-state measurement, their fields are overlaid. The azimuthal wavenumber of this reflected vortex, or its topological charge, is given by $k_\theta = l + 2m$ and is increased by twice the order of the structure, while its angular velocity $v_\theta = 2\pi/k_\theta$ is respectively decelerated.

So far, neither such a reflected vortex nor any other subsequent vortices have been experimentally observed. As we show in Fig. 3 (A, B), the predominantly explored slit geometry has poor reflectance (Fig. 3 B) of up to a maximum of 16%. Moreover, the slit reflectance maxima are not aligned with the coupling efficiency maxima in the structural parameter space and therefore slits do not perform as simultaneously good couplers and reflectors. Finally, in addition to the poor reflectance, plasmonic propagation losses further diminish the SPP power in the reflected plasmons by the time they reach the center to the formation of reflected vortices. The last two reasons and the lack of time resolution in previous experimental work practically eliminated the reflections from the experiments and solidified the notion of obtaining a single vortex from spiral-shaped vortex generators within the community.

In the following, we experimentally demonstrate that by using the time-resolved signal obtained by TR-PEEM, and a proper slit or ridge design, the generation of higher-order vortices (Fig. 4) by reflection can be enhanced and made detectable. For our study we use a 160nm gold layer sputtered onto a silicon substrate, and mill 500nm wide slits using a focused ion beam forming a plasmonic vortex generator of order *m=5*. The measurement was performed by time-resolved Photoemission Electron Microscopy (TR-PEEM). To detect the weak slit-reflected vortex, we employed the time-delay Fourier filtering technique of the TR-PEEM signal that increases the signal-to-noise of the desired contribution. In this process, the raw PEEM data is pixel-wise filtered in the time-delay conjugated domain to contain only the $\omega$ oscillating contributions, where $\omega = 2\pi c/\lambda_0$ is the frequency of light at $\lambda_0$=800nm [23].

Fig. 4 shows snapshots of the propagating vortex taken with the probe pulse at a time delay Δt after the launching of the vortex by the pump pulse at the boundaries. Note that due to the non-linear subtractive spin-orbit coupling between the circularly polarized light of the probe pulse and the propagating plasmonic vortex, inherent to the TR-PEEM imaging process [23], the number of lobes in the experimentally resolved image differs from the propagating plasmonic vortex (*l*) by – $\sigma$ *($l_{exp} = l - \sigma$)* that means the helicity of the probe pulse has to be subtracted. Fig. 4 (A) shows the first vortex of order 6 measured with right handed circularly polarized light ($\sigma = 1$) resulting in $l_{exp}$=6-1=5 azimuthal lobes. The snapshot is taken 58.5 femtoseconds after the launching of the plasmonic pulse by the boundaries. Fig. 4 (B) is taken 79 femtoseconds later, 137.5 femtoseconds after the launching of the pulse. It shows 15 azimuthal lobes corresponding to a plasmonic vortex of order $l = 16$. During 137.5 femtoseconds the plasmons propagate $d_{prop}$~41.25 microns and in a vortex generator with a radius of 12 microns, this corresponds to slightly more than one and a half round-trips within the structure. We thus observe a fully developed vortex of order $l = 16$ within the plasmonic vortex generator of order $l = 5$, corresponding to a vortex that is created due to the reflections of the plasmons from the slits defining the structure.

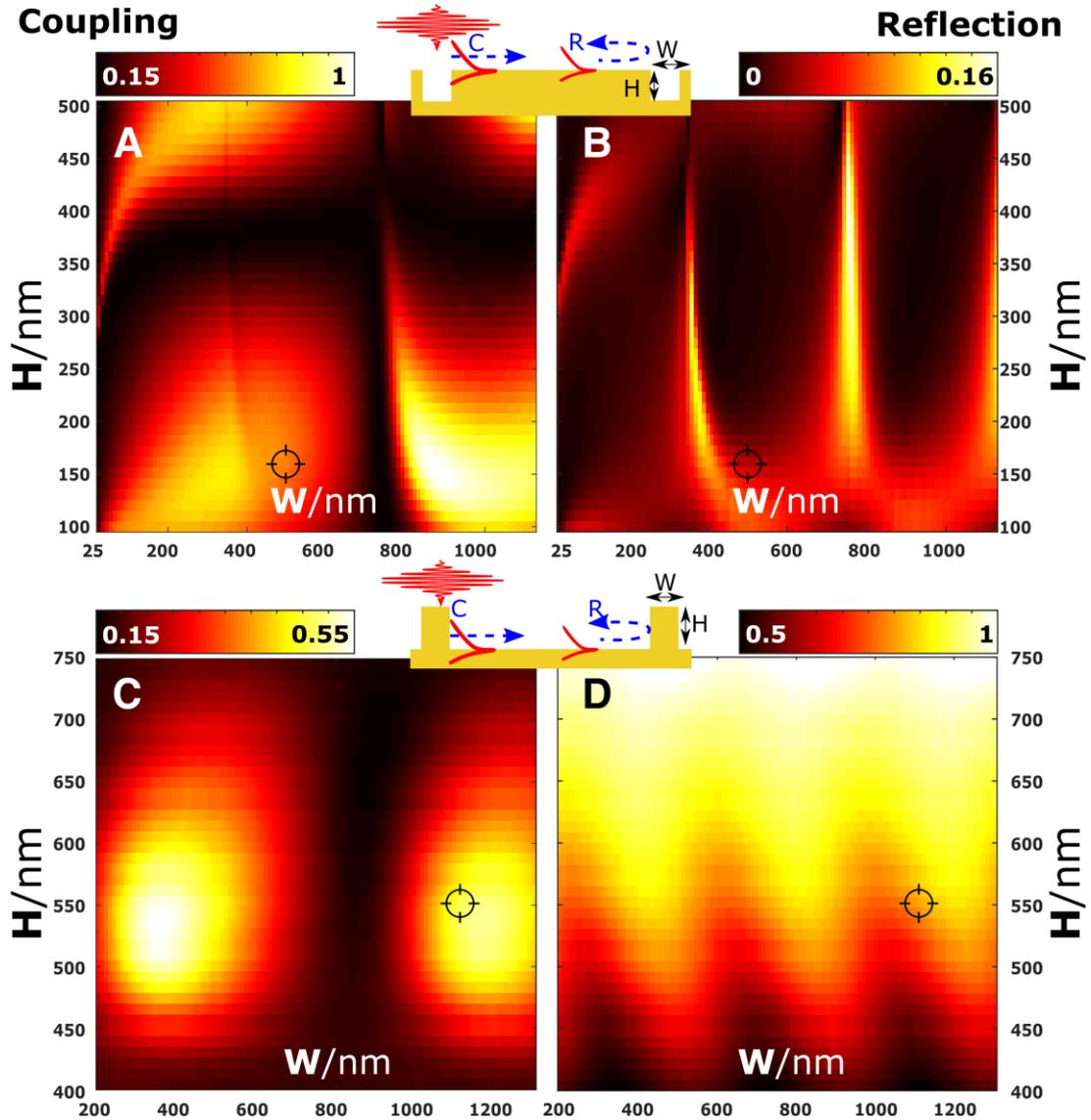

**Figure 3.** Finite Difference Time Domain calculations of coupling efficiency (A/C) and reflection (B/D) of a slit/ridge as a function of height and width. It can be seen that a ridge structure can function simultaneously as an efficient coupler and nearly perfect reflector whereas the slit cannot. The intersections show our fabricated structure having W=1110nm and H=550nm resulting in ~85% reflection at nearly maximal coupling strength accessible by the fabrication method.

The propagation loss associated with the distance traveled by the plasmons is given by $l_{prop}=\exp(-\Im(k_{spp}) \cdot d_{prop})$ where $\Im(k_{spp})$ is the imaginary part of the plasmonic wavenumber. The reflectance of the slit with our parameters shown in Fig. 3 (B) is $r_{slit}$=0.1. The slit-reflected vortex is therefore $l_{prop} \times$ (1-$r_{slit}$) times weaker than the original vortex, to the extent that it is not visible in the raw TR-PEEM signal.

To truly use and control the reflection-generated high order vortices, one must increase their relative intensity. We achieve that by augmenting the conventional plasmonic vortex generator and forming a plasmonic vortex cavity with boundaries consisting of golden walls (ridge structure) instead of slits. As shown in Fig. 3 (C, D), even though golden ridges couple 45% less energy from the impinging light pulse into the plasmons than slits do, they can achieve almost perfect reflectance. More importantly, as opposed to slits golden ridges can serve both as good couplers and nearly perfect reflectors as the two efficiency maxima are aligned. Our ridge design facilitates ~85% reflectivity for a wide range of incidence angles enabling the vortex generation within the cavity. This contrasts with Bragg reflectors that while achieving similar reflectivity using many grating periods [55], do so only for a narrow range of incident plasmonic wave vector angles.

To construct a ridge based plasmonic cavity we employ the process of template stripping (Fig. 1 D) [50,52–54]. We begin with patterning a single crystalline silicon substrate with the negative image of the desired cavity. After native oxide removal, we grow 50 nm thick thermal oxide layer and sputter a 700nm layer of gold without any adhesion layer. In the final stage, the top surface of the gold is glued onto a glass substrate and peeled off the silicon template (see full recipe in the SI). The resulting structure provides a flat golden surface and the desired ridges (Fig. 1 E).

The results in Fig. 1 (A-C) show a cavity of order 5 which is excited with right hand circularly polarized light ($\sigma = 1$) resulting in a vortex of order 6 (Fig. 1 A) and two subsequent vortices of orders $l=16$ and $l=26$ upon two consecutive reflections. The vortices show 5, 15, and 25 azimuthal lobes due to the nonlinear interaction with the probe pulse as discussed above and the radial distributions correspond well to the theory (see SI). The first and second reflections are imaged at subsequent time delays of 125.9 fs and 131.5 fs and correspond to ~37.77 and ~39.45 microns traveled by the plasmons. This distance is within the pulse duration correspondence to a round trip (from the center and back) within the cavity having $r_i=18$ µm. The limiting factor for observing the higher-order reflections is the maximal physical retardation of our delay stage.

We thus obtain theoretically and experimentally that for a continuous-wave experiment, the steady-state field distribution within plasmonic vortex generators and vortex cavities, in particular, is actually given by,

$$E \approx \sum_n a_n \cdot J_{l+2 \cdot n \cdot m}(k_{spp} r) \cdot e^{-i(l+2nm)\theta}$$

where $a_n$ is a proportionality coefficient convoluting the reflection coefficient and the losses accumulated by propagation within the cavity between each subsequent reflection. This is in contrast to the widely assumed single term $J_l(k_{spp}r) \cdot e^{-il\theta}$. Our finding shows that the $n^{th}$ generated vortex pulse carries OAM of the order $l_n = l + 2 \cdot n \cdot m$, indicating that each subsequent interaction with the boundaries of a vortex cavity of order $m$ decelerates the angular velocity by providing additional $2 \cdot m$ units of OAM.

From a physics standpoint, we have here an interesting cavity which has chiral mirrors that are nonlocal with respect to angular momentum. While regular mirrors of a cavity are local, reflecting the impinging field uniformly, reflection from a photonic crystal or a transverse grating is nonlocal, [56] transferring linear momentum to the reflected fields. Here we have a chiral mirror which is nonlocal with regards to angular momentum. The chirality breaks the symmetry so that any reflection always increases the angular momentum. This contrasts with reflection from typical gratings, which are nonlocal with respect to linear momentum and provide both an increase and decrease the linear momentum by the reciprocal vector of the grating.

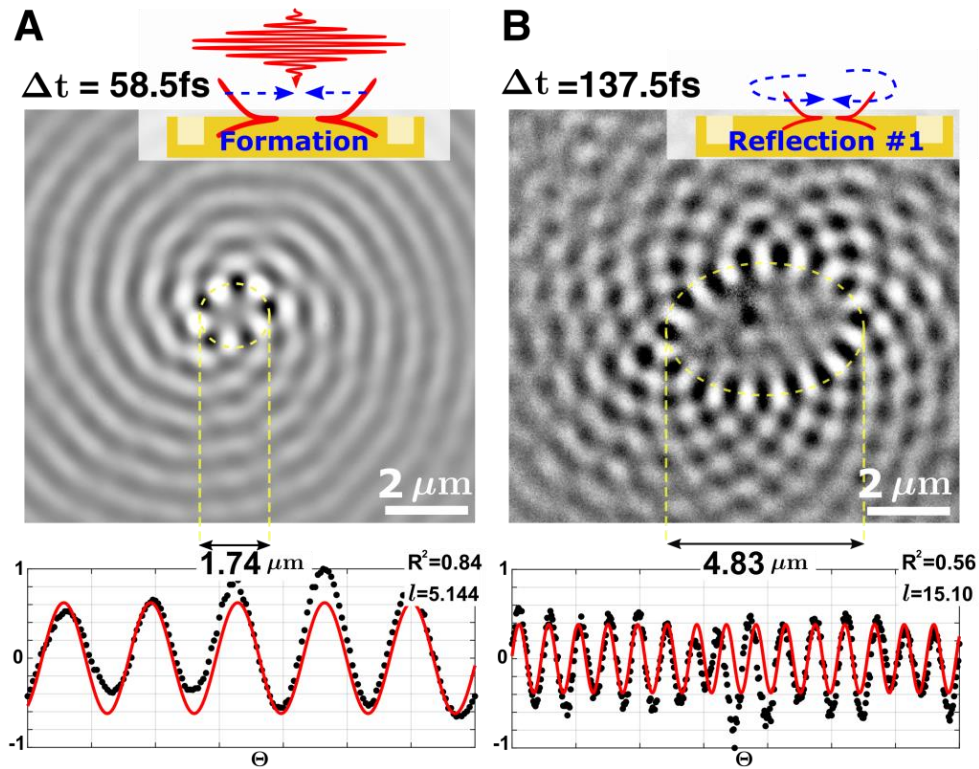

**Figure 4.** Experimental result of a cavity of order m=5 made with a slit reflector. (A) Formation stage of the initial vortex of order $l=5+1$ and the subsequent first (B) reflection of the wave packet, forming a vortex of order $l=15+1$.

The snapshots show 5 and 15 azimuthal lobes imprinted into the photoemitted electrons due to the nonlinear mechanism of our technique [23]. The bottom insets show azimuthal and radial fitting of the diameter as well as the number of lobes in the main vortex signal. The vortices are elliptically deformed due to a slight tilt misalignment of the ion beam stage during the sample fabrication (see SI).

Our results provide an experimental glance into a decelerating cavity, revealing how the boundary conditions of the cavity produce temporally increasing topological charge producing higher OAM modes with each reflection. By quantifying the rule governing the progression of the OAM pulse train within such cavities, we thus introduce a new degree of freedom – the reflection from structural boundaries – as a tool to tame surface-confined OAM. Besides the fundamental interest posed by non-local cavities and going beyond the paradigms established in the plasmonic community, we believe that plasmonic vortex cavities could be used to miniaturize time windowing techniques such as pump-probe-like quantum initialization schemes, the torque exerted by the vortex train may be used to increase the rotational power of plasmonic tweezers and enable the study of new phenomena as the break-down of plasmonic vortices when the reflection-generated vortices become larger than the cavity. By combining the presented cavities with a spatially distributed gain medium they could potentially be used as tunable generators of high order OAM. Finally, future studies of vortex lattice cavities could reveal how the cavity is populated with lattices having a temporally evolving topology.

## Acknowledgments

G. S. and M. O. acknowledge support from the Israeli Centers of Research Excellence "Circle of Light." We acknowledge the Micro-Nano Fabrication Unit (MNFU) Technion for support with sample fabrication, we acknowledge Dr. Guy Ankonina for the sputtering of gold, Dr. Jacob Schnider for help with thermal Oxide growing and Dr. Lior Gal for e-beam gold deposition. G.S. would like to acknowledge the support from the Clore Israel Foundation and the Schmidt Science Fellowship. We acknowledge the Nano Structuring Center Kaiserslautern for support with the sample fabrication. E.P. acknowledges support from the Max Planck Graduate Center with the Johannes Gutenberg University Mainz and the TU Kaiserslautern through a PhD fellowship.

# Supplementary Material for: Orbital angular momentum multiplication in plasmonic vortex cavities

G. Spektor[1,2,3], E. Prinz[4], M. Hartelt[4], A.K. Mahro[4], M. Aeschlimann[4] and M. Orenstein[1].

[1]Department of Electrical Engineering, Technion - Israel Institute of Technology, 32000 Haifa, Israel.
[2]Associate of the National Institute of Standards and Technology, Time and Frequency Division, Boulder, Colorado 80305, USA (curr. address)
[3]Department of Physics, University of Colorado, Boulder, Colorado 80309, USA (curr. address)
[4]Department of Physics and Research Center OPTIMAS, University of Kaiserslautern, Erwin Schroedinger Strasse 46, 67663 Kaiserslautern, Germany.

spektorg@campus.technion.ac.il

**Methods – Sample preparation for template stripping**

|  | **Processing** | **Container** | **Time** |
|---|---|---|---|
| 1 | HF - Remove the Oxide from wafer – Maximal HF concentration | Plastic | |
| 2 | Water | Plastic W/Holes | |
| 3 | Spin Dry | | |
| 4 | Piranha- $H_2SO_4$ : $H_2O_2$ - **2 : 1** | Glass | 10 min |
| 5 | Water | Plastic W/Holes | 7 min |
| 6 | HF **1 : 50** | Plastic | 10 sec |
| 7 | Water | Plastic W/Holes | 1 min |
| 8 | Hydrochloric - $H_2O$ + $H_2O_2$ + HCL , **6 : 1 : 1, Heat to 75°C** | Glass + Cover | 10 min |
| 9 | Water | Plastic W/Holes | 5 min |
| 10 | Spin Dry | | |
| 11 | Create structured patterns in silicon (Focused Ion Beam This work). | | |
| 12 | Grow 100-200nm of thermal oxide to prevent Gold-Silicon Adhesion. | | |
| 13 | HF **1:10** | Plastic | ~5-10sec |
| 14 | Water | | |
| 15 | AU deposition (Sputtering in this work) | | |
| 16 | Template Stripping | | |

The patterning in step 11 should take into account that while oxidizing a bare silicon surface (step 12), 46% of the oxide thickness will lie below the original surface, and 54% above it. Therefore, the width of the patterned silicon slits in step 11 will be altered upon oxidation and should be pre-modified at patterning to produce the desired dimensions.

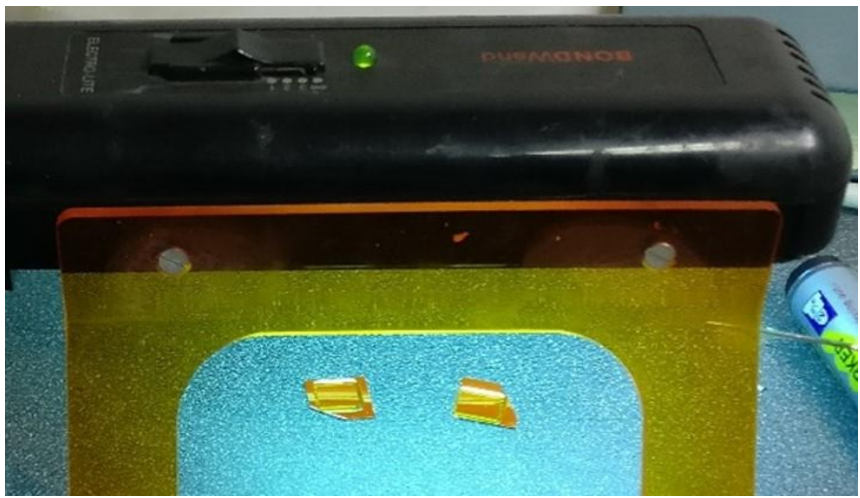

**Figure S1:** Example of exposing a UV-curable epoxy in order to glue a glass substrate onto the Silicon piece sputtered with gold, before stripping the gold with the glass substrate.

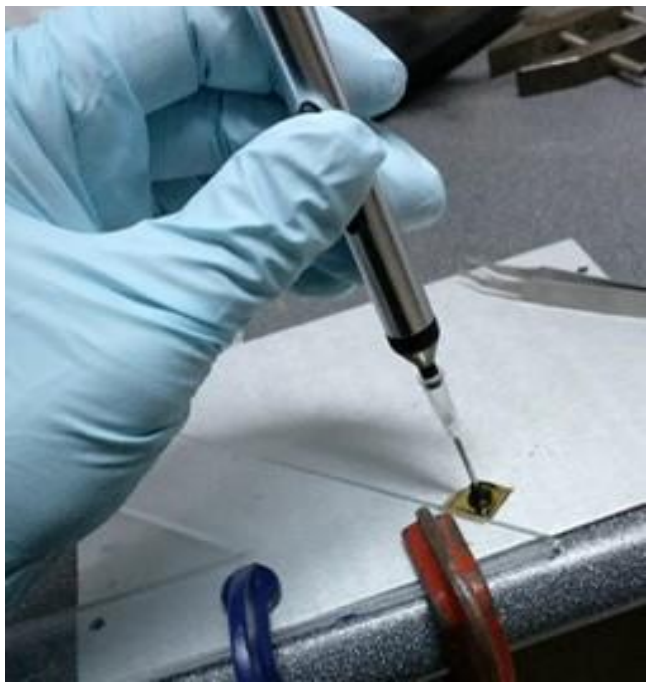

**Figure S2:** Example of stripping the glass substrate with the gold using a suction holder.

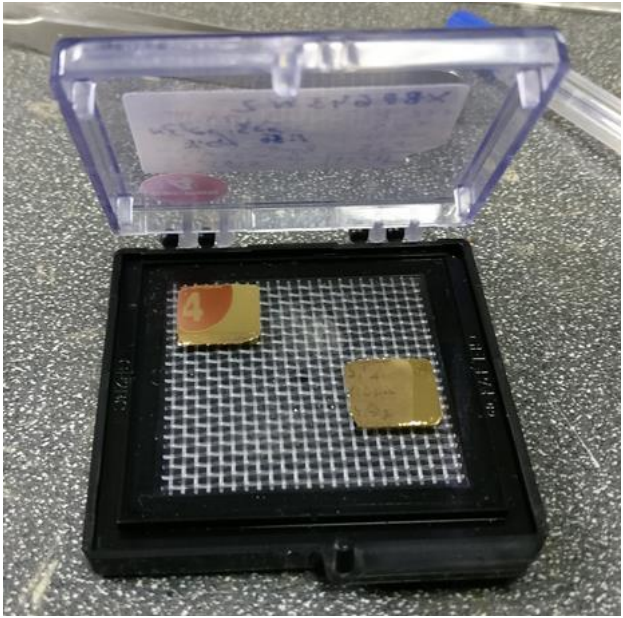

**Figure S3:** Finished template stripped samples.

## Bessel Functions

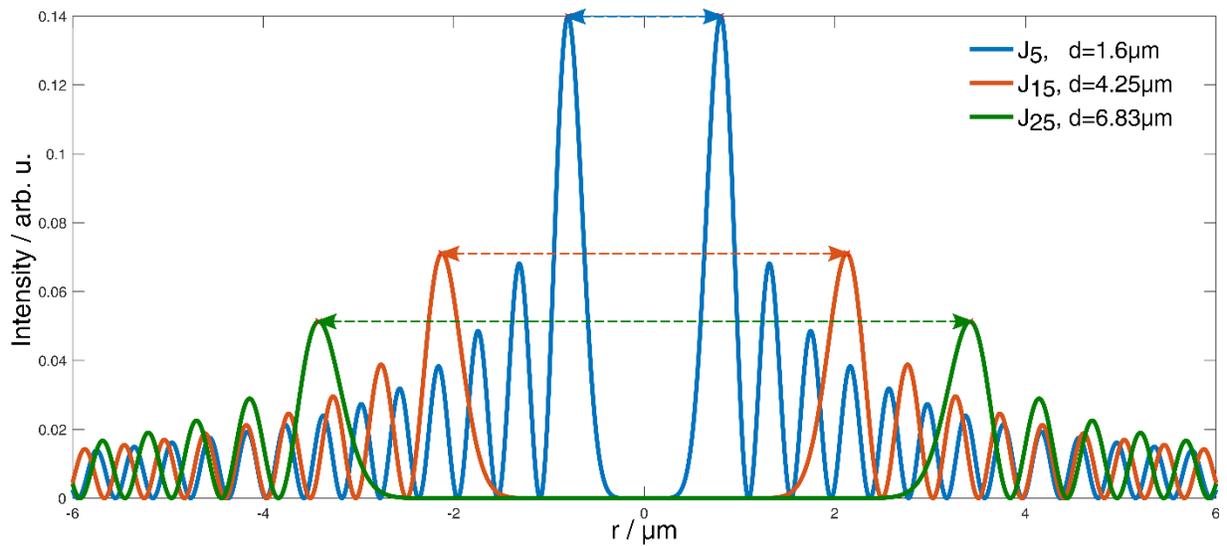

**Figure S4:** Different orders of Bessel functions, calculated via $|J_m(k_{spp}r)|$ with $\lambda_{spp} = 784$ nm and $m = 5, 15, 25$, corresponding to the radial profiles of the plasmonic vortices shown in the main manuscript (Fig. 1 and Fig. 4). One can see the increase of main lobe diameter with increasing Bessel order and the intensity of the main lobe decreases.

## Slightly deformed slit cavity

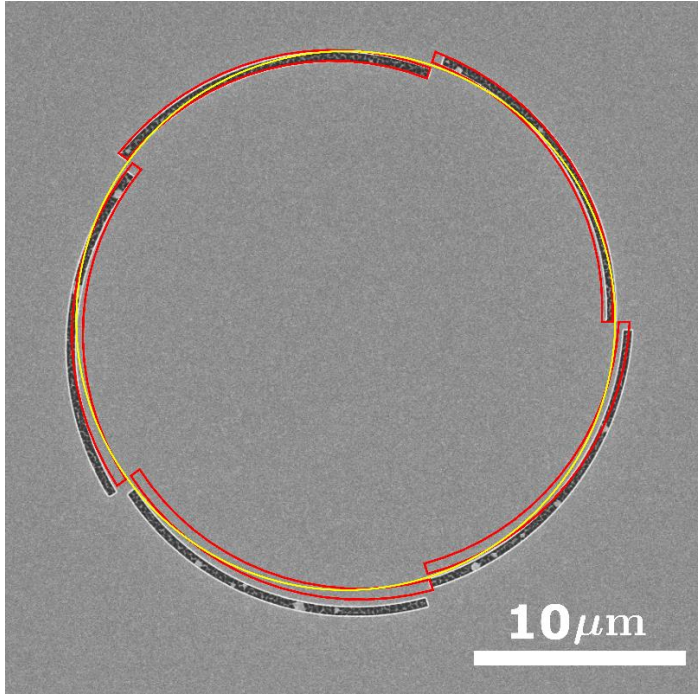

**Figure S5:** Scanning electron microscope (SEM) image of a plasmonic cavity with a slit reflector. Due to a slight tilt misalignment of the ion beam stage during fabrication it is deformed, leading to the formation of elliptical vortices (see Fig. 4 in the main manuscript). The yellow overlay is a circle and the red overlay shows the shape that the cavity was supposed to have.

**Movies:**

FDTD Simulations (performed with *Lumerical FDTD Solutions*):

- Reflection from a slit and from a ridge, same contrast: [Simulations_Reflection.mp4]

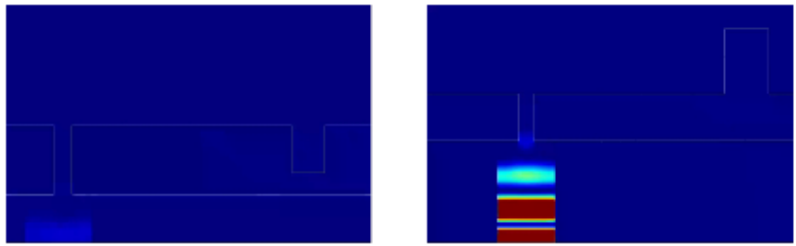

Experimental:

- Fourier-filtered reflections in ridge cavity (initial vortex + first & second reflection): [ridgeData.mp4]

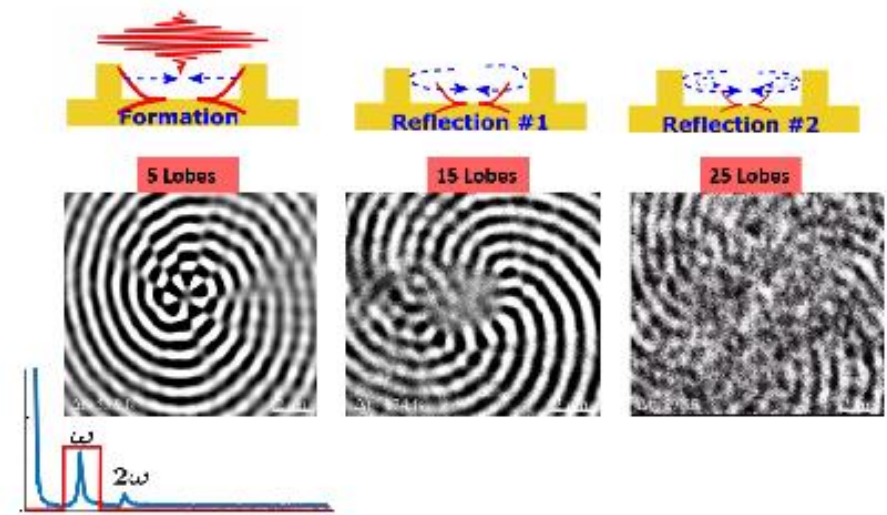

- Fourier-filtered reflections in slit cavity (initial vortex + first reflection): [slitData.mp4]

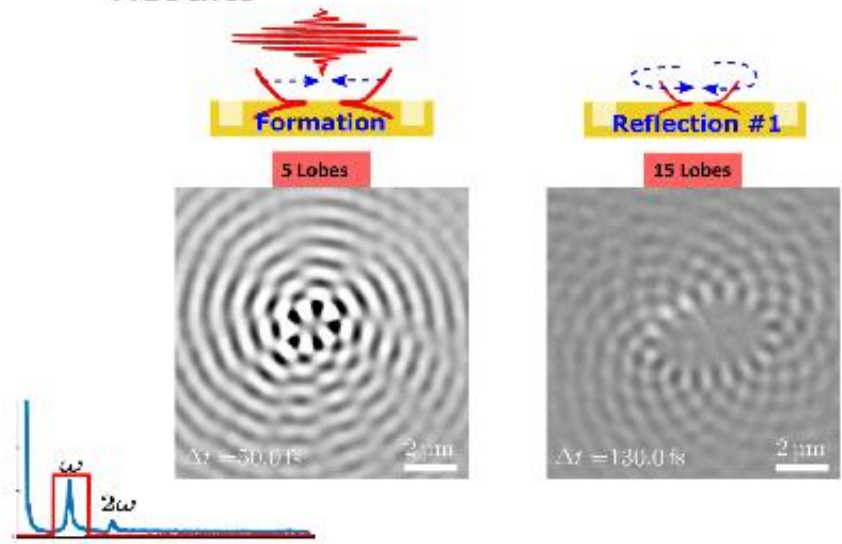

Spektor et al., PRX 9, 021031, 2019